\documentclass[lettersize,journal]{IEEEtran}

\usepackage{blindtext, graphicx}
\usepackage{csquotes}

\usepackage{cite}

\usepackage[cmex10]{amsmath}
\usepackage{amsthm}
\usepackage{tensor}
\usepackage{optidef}
\usepackage{array}
\usepackage{mathrsfs}
\usepackage[tight,footnotesize]{subfigure}
\usepackage{stfloats}
\usepackage{float}
\def\*#1{\mathbf{#1}}

\newtheorem{proposition}{Proposition}
\usepackage{xcolor}
\definecolor{light-gray}{gray}{0.5}

\newcommand{\REV}[1]{\textcolor{black}{#1}}
\usepackage{amsfonts}

\begin{document}
%
\title{\REV{Deterministic Patterns for Multiple Access with Latency and Reliability Guarantees}}

\author{Rados\l{}aw Kotaba, 
        Roope Vehkalahti, 
        \v{C}edomir Stefanovi\'{c},~\IEEEmembership{Senior Member,~IEEE},
        Olav Tirkkonen,~\IEEEmembership{Fellow,~IEEE}
        and~Petar Popovski,~\IEEEmembership{Fellow,~IEEE}%
        
\thanks{Manuscript received March 13, 2022. This paper was supported in part by the Villum Investigator Grant "WATER" from the Velux Foundations, Denmark, and by Business Finland under the project "Extreme Machine Type Communications for 6G".
}
\thanks{R. Kotaba, \v C. Stefanovi\'{c} and P. Popovski are with the Department of Electronic systems, Aalborg University, 
Denmark  (e-mail: \{rak, cs, petarp\}@es.aau.dk).}
\thanks{R. Vehkalahti is with the Department of Mathematics and Statistics, University of Jyv\"{a}skyl\"{a}, 
Finland   (e-mail: roope.i.vehkalahti@jyu.fi ).}
\thanks{O. Tirkkonen is with the Department of Communications and Networking, Aalto University, 
Finland
(e-mail: olav.tirkkonen@aalto.fi).}
}


\maketitle

\begin{abstract}
\REV{We study a scenario in which multiple uncoordinated devices aim to achieve reliable transmissions within a given time frame. The devices are intermittently active and access a shared pool of channel resources in a grant-free manner by utilizing multiple transmissions (\textit{K}-repetition coding). This allows them to achieve diversity and improve the reliability within a certain latency constraint. We focus on two access methods: one where devices choose \textit{K} slots at random and another one where the \textit{access patterns} are deterministic and follow a specific code design, namely the Steiner System. 
We analyze the problem under two signal models that involve different complexity for the receiver.
\emph{First}, collision model is considered, where only interference-free transmissions can be used and combined. \emph{Second}, a model treating interference as noise is analyzed, where the receiver is capable of utilizing all \textit{K} replicas, applying maximum ratio combining (MRC).
For both signal models, we investigate receivers with and without successive interference cancellation (SIC). We develop approximations and bounds for the outage probabilities that very closely match simulation results. Overall, we show that deterministic access patterns have the potential to significantly outperform random selection in terms of reliability. Furthermore, deterministic access patterns offer a simplified system design.}

\end{abstract}

\begin{IEEEkeywords}
grant-free, radio resource management, reliability-latency constrained communications
\end{IEEEkeywords}

\IEEEpeerreviewmaketitle

\section{Introduction}\label{sec:intro}

The latest generation of wireless systems, 5G networks, are becoming widely adopted~\cite{5g_report}, while researchers and the industry already plan their next steps by laying ground for 
forthcoming 6G technologies~\cite{Uusitalo2021}. 
\REV{Importantly, the shift to 5G and beyond is not just about the need for higher data rates, but also an increasing number of applications that require communication reliability within a certain constraint on latency, such as the ones involving tactile interaction, intelligent transportation and factory automation~\cite{5gusecase,5g_ngmn}.
At the high end of the spectrum of such use cases are the ones requiring ultra-reliable low-latency communications (URLLC), which are characterized by 
stringent end-to-end (E2E) latency 
and reliability constraints,  
with the probability of successful delivery of a packet at least 
$99.999\%$ within a latency budget between $0.5-2$ ms~\cite{5gusecase}.  There are also a number of use cases with less stringent latency and reliability requirements~\cite{popovski2022perspective}, while 6G research foresees even more tight requirements for mission critical subnetworks~\cite{Uusitalo2021}.}

The primary challenge in designing systems with latency and reliability guarantees is 
not to overly compromise 
spectral efficiency in the process~\cite{tradeoff_thesis}.
This is particularly difficult to achieve in the uplink, which in 
conventional cellular networks is centrally managed by the base station (BS) and relies on either explicit grants,
incurring high latency due to 
signalling,
or pre-allocation of resources
resulting in low latency but being inefficient for intermittent traffic.
Clearly, new uplink access protocols and modes of operation have to be devised for 
6G in order to fulfill the demanding latency and reliability targets.

One way to tackle the problem is to rely on random access communication protocols.
In grant-free (GF) access~\cite{grantfree:berardinelli}, 
user equipments (UEs) are allowed to transmit data without prior, explicit scheduling. 
A certain portion of bandwidth is 
provided to a group of UEs who can use it whenever they have data to send.
The benefit is a significant reduction in the signalling overhead and latency.
As a matter of fact, scheduling contributes the most to 
E2E delay, making it the main bottleneck in URLLC system design~\cite{scheduling_delay}.
 GF is a particularly suitable solution for 
traffic that is relatively infrequent and irregular~\cite{grant_free_survey}.
Due to the sharing of 
resources and lack of coordination, GF is inherently less reliable than its grant-based counterparts, 
as the uplink signals of the transmitting users are prone to collisions and interference. 
To compensate for this, additional mechanisms 
to improve the reliability of GF access are needed.

\REV{In this work, we study 
GF multiple access in which users apply access patterns, i.e. sequences consisting of multiple redundant transmissions, to achieve higher communication reliability. Our framework follows the line of work started with CRDSA/IRSA in which each randomly accessing device sends multiple packet replicas \cite{crdsa1,irsa1}, rather than a single one as in plain ALOHA. The model involves a shared pool of resources - a short, periodic frame composed of limited number of slots, that makes our contribution relevant in scenarios with tight latency constrains. 
We focus on 
comparing fully random selection of slots to 
methods where deterministic patterns are pre-assigned to users. Specifically, we consider deterministic patterns arising from 
\emph{Steiner systems}. 
These have the desirable feature that their construction ensures that two patterns can share at most $t-1$ slots, where $t$ is a design parameter,
thus providing guarantees in terms of the 
number of collisions/interference.}


\REV{We 
analyse the performance of grant-free access 
systems in terms of its outage probability and spectral efficiency, considering two different signal models.
The first one resembles a traditional slotted ALOHA system, where collisions are \textit{destructive}, i.e. only the slots that contain a transmission of a single device can be used.
Unlike 
slotted ALOHA,
we allow the receiver to combine multiple collision-free replicas from a given user.
In the second model, the receiver is capable of utilizing all transmissions and performs maximum ratio combining (MRC) accounting for different SINRs.
Furthermore, for each of the two signal models we consider
receivers with and without successive interference cancellation (SIC). 
}

\REV{
A particularly important contribution are the approximations and bounds developed for the collision model with SIC, and for 
MRC without SIC, which 
are two analytically challenging cases.
The developed expressions match very closely the extensive simulation results obtained with Monte Carlo methods; these approximations are especially valuable in the context of dimensioning systems with strict latency and reliability guarentees, where relying on simulations alone is often not feasible due to the sheer number of samples required.
To the best of the 
our knowledge, similar results have not been reported before. 
}

\REV{
The most important engineering insight of this paper is that in all of the aforementioned configurations, access patterns given by a Steiner system exhibit clear gains over their 
random counterparts.
Furthermore, their regular structure simplifies the overall system design.  
As such, we believe that Steiner systems make for a compelling solution in the design of grant-free access.}


Finally, we note that this work extends our prior contribution \cite{boyd:spawc} in several meaningful ways. 
Firstly, we provide an in-depth analysis and develop approximations and bounds that go well beyond the results reported earlier.
We also present formal proofs of the combinatorial results in \cite{boyd:spawc} that treat the distribution of the number of collision-free slots and number of interferers, and which were omitted due to space constraints.
Secondly, we extend the scenario by considering receiver with SIC capabilities.
We also broaden the scope by considering other Steiner systems with different parameters (frame length and number of repetitions).
Lastly, we discuss the limitations of access methods based on random selection highlighting issues with their practical implementation.

The rest of the paper is organized as follows. 
A brief account of the preliminaries and the related work is presented in Section~\ref{sec:intro:prior}.
The system and signal model are presented in Section~\ref{sec:sys_model}.
In Section~\ref{sec:patterns} we introduce the two types of access patterns and discuss their properties.
Then, in Section~\ref{sec:receiver} we consider different receiver processing techniques and provide their thorough analysis in the context of the access patterns from Section~\ref{sec:patterns}. This is complemented by both analytical results and corresponding simulations.
In Section~\ref{sec:sys_design:random}, we discuss the deficiencies of the Random selection approach and the challenges wrt. its practical implementation. 
In Section~\ref{sec:sys_design:resources} we compare different Steiner Systems using the analytical results derived earlier.
Lastly, in Section \ref{sec:concl} we offer final conclusions that close the paper.

\section{\REV{Preliminaries and Related Work}}
\label{sec:intro:prior}

The simplest and at the same time most effective solution to increase the reliability of a random access scheme is to introduce redundancy through multiple transmissions~\cite{dsa}.
The use of such strategy increases the chances that at least some of the replicas reach the BS uninterfered, and secondly, by providing diversity 
that allows to combat the negative effects of the fading channel.
In a typical approach, 
there is a pool of resources and
the UEs decide randomly in which resources 
to perform multiple transmissions. 
However, the transmissions of the individual users can be structured into \textit{access patterns}.
The patterns can be constructed in many ways and with different goals in mind, but in general they aim to provide certain reliability guarantees~\cite{Boyd2018}. 
The drawback of such solution is that their assignment requires 
coordination with the BS and signalling, which makes it less flexible than purely random selection.
However, this operation can be integrated into the registration procedure that each device has to perform anyway when it 
attaches to the BS.
Furthermore, even in a fully random scheme the device needs to be configured at least with the number of repetitions and portion of bandwidth where the grant-free pool is located. 

On the receiver side, there is a possibility to implement 
successive 
interference cancellation. 
With SIC, it is possible to iteratively decode signals by gradually removing the interference.
In each round the packets that were successfully decoded in the preceding rounds can be subtracted 
from the received signal, thus improving the signal-to-interference-plus-noise (SINR) of the remaining ones.
This is 
especially relevant when dealing with traffic that is non-orthogonal by design~\cite{crdsa1}. 

Fundamentally, the GF techniques descend from one of the most well known concepts in the field of communication -  ALOHA\cite{slottedAloha}.
Since its inception, many extensions have been proposed. One of them is the Content Resolution Diversity Slotted ALOHA \cite{crdsa1,crdsa2}, which utilizes multiple transmissions (with the goal of achieving diversity) and the interference cancellation. 
In \cite{irsa1}, authors analyze a variant of this scheme - Irregular Repetition Slotted ALOHA (IRSA) in which the number of repetitions is not fixed, but follows a certain discrete distribution. Furthermore, in \cite{irsa2} the analysis is extended to the Rayleigh fading channel and optimization of the repetition degree is presented.
Another extension coined Coded Slotted ALOHA (CSA)\cite{csa1} involves transmitting different coded version of the packet (redundancy versions) rather than exact replicas, which allows to achieve better granularity in terms of transmission rate. In \cite{cra2} the author analyzes the throughput of CSA in a multichannel Rayleigh fading scenario.
In these works, the primary focus is on the maximization of the load of the system (in terms of the number of transmitted packets per slot) for which the successful decoding probability of a packet tends to 1. 

GF access methods 
designed specifically for 5G URLLC were 
researched in \cite{abreuThesis}.
The author investigates different repetition and retransmission schemes in realistic scenarios based on a detailed system level simulator. 
In \cite{grantfree:capacity}, the authors focus on the combinatorial aspects of a repetition-based 5G GF scheme, namely the probability of collisions, and evaluate achievable reliability and latency levels as a function of the number of UEs, amount of pre-allocated resources and number of replicas.
The GF repetition coding, its proactive version (where the UEs have the possibility of early termination), and the more traditional Hybrid Automatic Repeat Request (HARQ) based on feedback and retransmissions have been jointly evaluated and compared in \cite{harqGF}.
In \cite{choi_network} the author extends the idea of repetition-based schemes towards network coding, making it better suited for scenarios where devices have more than one packet to transmit at a time.
In the approach considered in \cite{sensingGF}, the resources are first assigned to a group of users based on sensing, and then the UEs avoid the collisions by signalling transmission announcements between themselves.

In addition to fully random GF schemes, 
pre-allocated access patterns and their design have been considered in the literature.
In \cite{comb1}, 
patterns based on a combinatorial designs were discussed.
However, the authors do not consider SIC and treat all the collisions as destructive. 
More recently, 
designs oriented towards interference cancellation have been considered in \cite{combBoyd},\cite{boyd:spawc} and \cite{combldpc}. 
In \cite{combBoyd} the patterns are constructed based on so-called locally thin codes, in \cite{boyd:spawc} they are based on the Steiner Systems, while in \cite{combldpc} authors use LDPC codes.
The idea to use deterministic access patterns also appeared in \cite{mimoPatterns}, where they are used in conjunction with multiple antenna processing at the BS.
Channel resources are divided into high and low contention parts over which power optimization is 
performed. 
The work in \cite{symbPattern} differs in that the patterns are applied on the symbol-level rather than over slots.
	

\section{System model}\label{sec:sys_model}

\begin{figure}[t!]	
	\centering
	\includegraphics[width=0.65\linewidth]{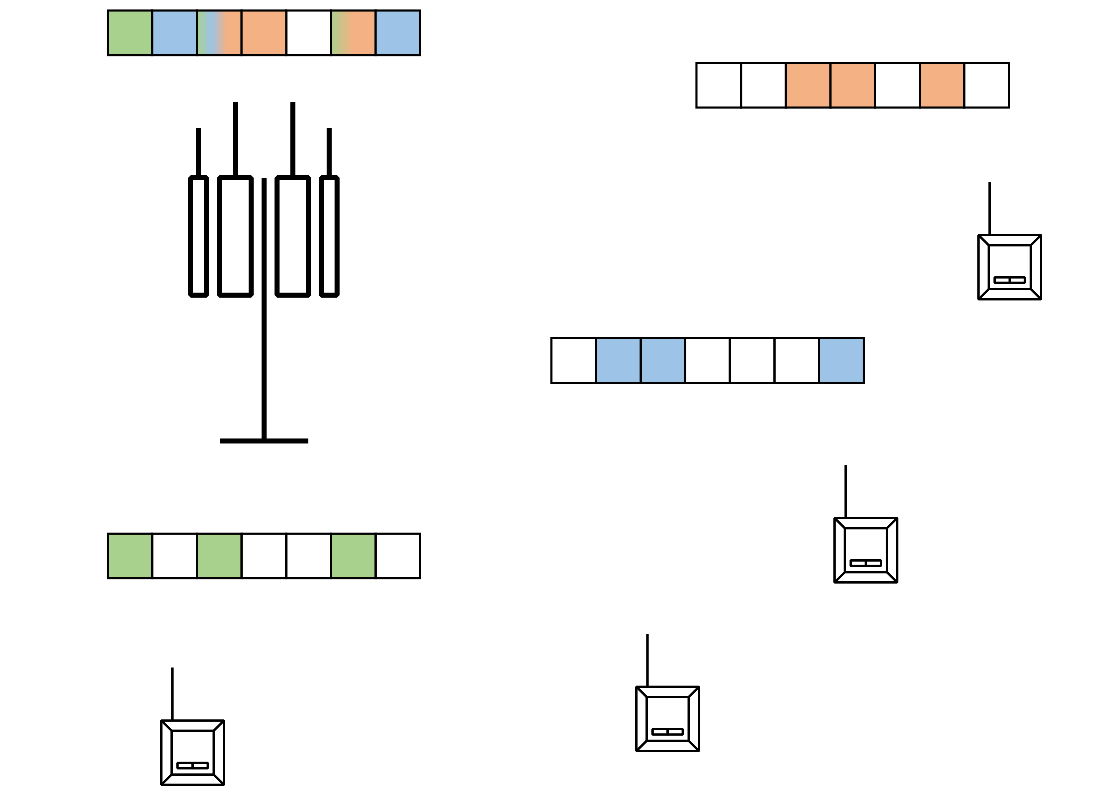}
	\caption{Example of the uplink access scenario with $K=3$ multiple transmissions over a frame of $M=7$ slots. There are $N=4$ UEs out of which $U=3$ happen to be active. Their transmissions cause collisions in slots $3$ and $6$.}
	\label{fig:system model}
\end{figure}

We consider a communication system with a single base station (BS) serving a population of $N$ intermittently active users (UEs) transmitting in the uplink. 
The shared wireless channel is composed of periodic frames, 
which are further broken down into $M$ 
slots\footnote{
The slots can be arranged in time, or they may 
represent different frequencies/groups of frequencies (subcarriers) or be 2-dimensional constructs similar to Resource Blocks in LTE/5G.}. 
We assume that UEs are independently activated in a frame with probability $b$, such that the total number of active devices $U$ follows a binomial distribution $f_\text{bin}(u;b,N)=\binom{N}{u}b^u(1-b)^{N-u}$.
Whenever active, a user selects $K$ out of $M$ slots in a frame and uses them to transmit its packet, thus employing 
$K$-repetition coding.
It is further assumed that all UEs transmit with the same rate\footnote{\REV{The assumption of equal transmission rates is common in the literature investigating slotted ALOHA access and makes the analysis tractable. Furthermore, the value of $R$ is typically low, entailing relatively small packets coded at a very low rate, which is in line with the chosen high-reliability use case considered in this paper.}} $R$ measured in bits per channel use (c.u.).
The described model is visualized in the example in Fig.~\ref{fig:system model}.
In general, $M$ is determined by the allowed latency \REV{and available bandwidth}, while $K$ is a design parameter that depends on the number of users $N$ and the activation probability $b$.

\REV{We consider a Rayleigh block fading channel, where the channel coefficients are independent across slots and UEs. 
In a frame there are $U$ active users forming a set ${\cal U}$.  Each  $u\in {\cal U}$ transmits a message $\*x_u$ consisting of $J$ complex modulation symbols and repeats it in $K$ slots. The transmitted symbols of the $U$ users are represented by the matrix $\*X \in \mathbb{C}^{U\times J}$
and are normalized such that $\mathrm{E}\left[ \left| x_{uj} \right| ^2 \right] = 1$.}

\REV{
At the receiver, the discrete time baseband representation of the channel output in a frame\footnote{We assume that transmissions are contained in a single frame, i.e., UEs are not supposed to transmit replicas of their packets over multiple frames as that would violate the implicit latency constraint.} is
\begin{align}
    \*Y & =  
      \left(\*G \odot  \*V \right) \*P \, \*X+\*W
    \label{eq:baseband}
\end{align}
where $\*Y \in \mathbb{C}^{M\times J}$ is the received signal consisting of $J$ samples for each of the $M$ slots.  
The channels of the active users are represented by the matrix 
$\*G \in \mathbb{C}^{M \times U}$  with zero-mean, unit-variance, circularly-symmetric complex Gaussian entries, modeling the underlying uncorrelated Rayleigh flat fading channels between the $U$ users and the base station in each of the $M$ slots. 
$\*V \in \{0,1\}^{M \times U}$ is s matrix representing the access patterns of the active users such that
\begin{align}
V_{m,u} = \begin{cases}
    1 & \text{if user $u$ transmitted in slot $m$} \\
    0 & \text{otherwise}\,,
\end{cases}
\end{align}
The channel and access pattern matrices are combined with the entry-wise (Hadamard) product $\odot$. 
Finally, 
$\*P = \text{diag} \left( (P_{1})^{1/2}, \dots, (P_{U})^{1/2} \right)$ is a diagonal matrix of square roots of average received powers, i.e. signal amplitudes of the active UEs,\footnote{\REV{Note that a  distance-dependent path loss term is absent in the signal model. Throughout this work we assume that UEs know the long-term statistics of the channel and based on that compensate for path loss accordingly. With $d_n$ the path loss of user $n$, the actual transmit power would be $d_n^{-1} P_{x}$.
This assumption allows us 
to omit the path loss term altogether.}}
and $\*W \in \mathbb{C}^{M\times J}$ models additive white Gaussian noise with zero mean and variance $\sigma^2$. 
}

\REV{
In the remainder of this work we assume that all active UEs 
have  the same average received power $P_x$. Consequently, 
the average received SNR is
\begin{equation}
\theta = \frac{P_x}{\sigma^2}.    
\end{equation}}

\REV{
The set of indices $\{ m: V_{m,u} = 1\}$ represent the access pattern of user $u$, i.e., the choice of slots which it used for transmission.
The users may follow access patterns that were pre-assigned to them, i.e., the columns in $\*V$ may be user specifc and fixed, or they may use random access patterns. Since each user uses only $K$ out of  $M$ available slots, we have 
$\sum_{m=1}^{M} V_{m,u} = K$, $u = 1, \dots, U$. 
}

\section{Access Patterns: Random vs. Deterministic}\label{sec:patterns}

In this section we introduce
two access methods that can be employed by the devices that try to communicate over a shared pool of resources modeled as slots, 
one relying on random selection, 
the other 
on pre-assigned, deterministic access patterns for users.

\subsection{Random selection}

We start with an approach in which users transmit their $K$ packet replicas over the $M$ available slots in a frame by selecting slots uniformly at random.
We are interested in determining the probability that a certain number out of $K$ transmissions of some user in the frame does not experience collisions with transmissions of the other users.

Consider a frame in which $U\geq 1$ out of a population of $N$ users is active and, without loss of generality, focus on a single, arbitrary user $u$. When there are $U$ users, a collision-free resource of a users is a slot where  none of the other $U-1$ users causes a collision. We have:
\begin{proposition}\label{proposition:free:random}
When $U$ devices employ random access patterns with $K$ repetitions in $M$ slots, the probability that an arbitrary user has $K'$ out of $K$ collision-free slots is 
\begin{equation}
p_{\mathrm{CF,R}} (K'|U) = \binom{K}{K'} \sum_{n=0}^{K-K'} (-1)^n a_n T_n,
\label{eq:collfree:random}
\end{equation}
where $a_n = \binom{K-K'}{n}$ and $T_n = \left(\binom{M-K'-n}{K}/\binom{M}{K}\right)^{U-1}$.
\end{proposition}

The proof 
can be found in Appendix~\ref{sec:appendix:proposition1}.

Another relevant metric 
that may have an impact on the decoding is the distribution of the number of interferers $L$ in a given slot in which the reference user $u$ is active.
Since users are free to select any of the $\binom{M}{K}$ possible random access patterns and do so independently from each other (with replacement), the probability that $L$ out of potentially $U-1$ interfering users is active in any of $K$ slots in which $u$ is active can be computed as
\begin{equation}
	p_{\mathrm{I,R}}(L|U) = \frac{ \binom{M-1}{K}^{U-1-L} \binom{M-1}{K-1}^{L} \binom{U-1}{L}}{\binom{M}{K}^{U-1}}  ~. 
	\label{eq:interferers:random}
\end{equation}
where $L \in [0,U-1]$.

\subsection{Deterministic patterns}

Another approach 
for contention-based access over a shared pool of resources is the one in which UEs have fixed, pre-assigned access patterns.
Such a solution is 
less flexible, as it requires 
coordination with the BS, who is responsible for assigning the patterns. 
It has, however, 
potential to greatly improve the overall reliability of the system.
Typically, a pattern would be assigned when the device registers with the BS for the first time or wakes up and re-synchronizes after being in power-efficient mode.
The patterns can be also periodically updated.
This may happen, for example, when the user population size changes and the resource pool needs to be adjusted; however such updates will occur relatively infrequently compared to the duration of the frame.

In this work we 
focus on the patterns that are given by a Steiner system
block design.
A Steiner system $S(t,K,M)$ can be considered as a $M$-dimensional constant-weight code, where each codeword has $K$ ones, and for any two distinct codewords $s_i,s_j \in S(t,K,M)$, $s_i \neq s_j$, 
the Hamming distance $d(s_i, s_j) \geq 2K -2(t-1)$.
In our case, the codewords represent access patterns of users, and 
the patterns of two users can collide on at most $t-1$ positions. 
%

The number of codewords in a Steiner-code is 
\begin{equation}
    C=|S(t,K,M)|=\binom{M}{t} \left/  \binom{K}{t}\right. 
    = \frac{M! \,(K-t)!}{K!\,(M-t)!}   \,.
\end{equation}
For fixed $K$ and $M$, the lower the $t$ is, the smaller is the
number of patterns in the codebook, and thus the number of supportable users. Since in this work our focus is on reliability and latency constrained applications, we limit our considerations to the case $t=2$ as it provides high reliability and the support for a massive number of devices in not required\footnote{Technically, the highest reliability is provided when $t=1$, however such case is trivial as it corresponds to the fully orthogonal allocation of resources.}.

Another property of Steiner systems is that their structure guarantees that the number of  patterns overlapping in any given slot is 
\begin{equation}
D=\binom{M-1}{t-1} \left / \binom{K-1}{t-1}\right.
 = \frac{K}{M}\, C
\,.    
\end{equation}
Thus in the worst case, when all 
$D$ access patterns which include a certain slot
are active, there are 
$D$ mutually interfering users in a slot.
As elaborated later, this is an important feature, as it allows to dedicate 
the right amount of resources for 
pilot sequences to ensure that no pilot collisions occur.

Analogously to the random selection, we have the following results for the Steiner system.
\begin{proposition}
When a random set of $U$ out of $C$ devices employ distinct access patterns from a $S(t,K,M)$ Steiner system, the probability that an arbitrary user has $K'$ out of $K$ collision-free slots is
\begin{equation}
	p_{\mathrm{CF,S}}(K'|U) = \binom{K}{K-K'} \sum_{n=0}^{K-K'} (-1)^n a_n W_n
	\label{eq:collfree:steiner}
\end{equation}
where $W_n = \binom{(C-1)-(D-1)(n + K')}{U-1}/\binom{C-1}{U-1}$. 
\label{proposition:free:steiner}
\end{proposition}

The proof is provided in Appendix~\ref{sec:appendix:proposition2}.

In terms of the number of interferers $L$, the analysis is straightforward.
In a slot in which an arbitrary user is active, there are only $D-1$ other patterns that could cause a collision and the selection is done without replacement due to the unique preassignments.
Hence, the probability that $L$ out of $U-1$ devices select one of them while the rest of the devices select any of the remaining $C-D$ patterns is:
\begin{equation}
	p_{\mathrm{I,S}}(L|U) =  \frac{\binom{D-1}{L} \binom{C-D}{U-1-L}}{\binom{C-1}{U-1}}~.  
	\label{eq:interferers:steiner}
\end{equation}

\begin{figure*}[t!]%
\centering
\subfigure[]{
\label{fig:pk}%
\includegraphics[width=0.325\textwidth]{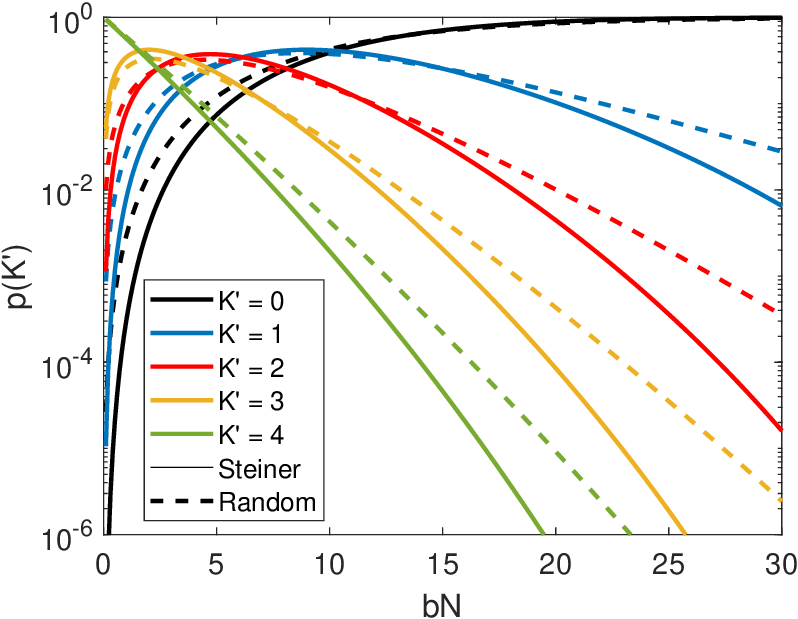}}%
\hfill
\subfigure[]{
\label{fig:pl}%
\includegraphics[width=0.325\textwidth]{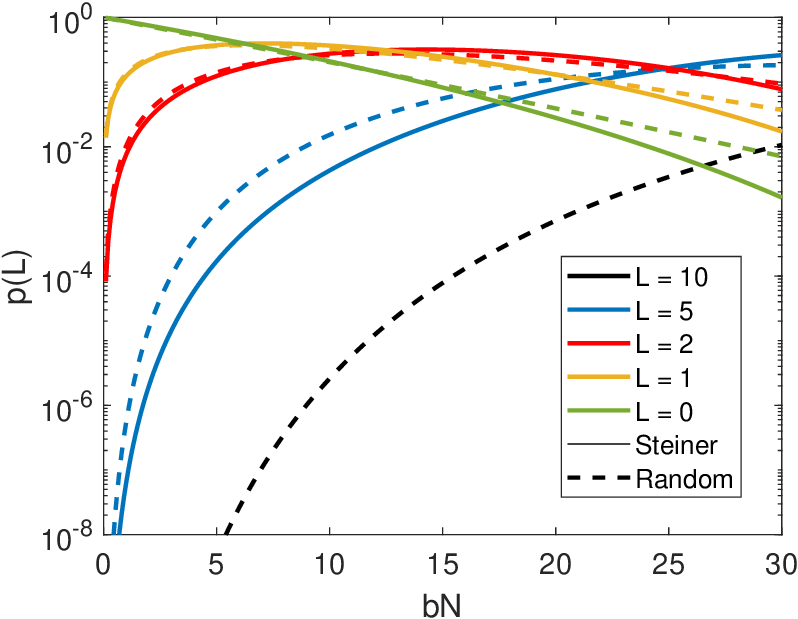}}%
\hfill
\subfigure[]{
\label{fig:cdfl}%
\includegraphics[width=0.315\textwidth]{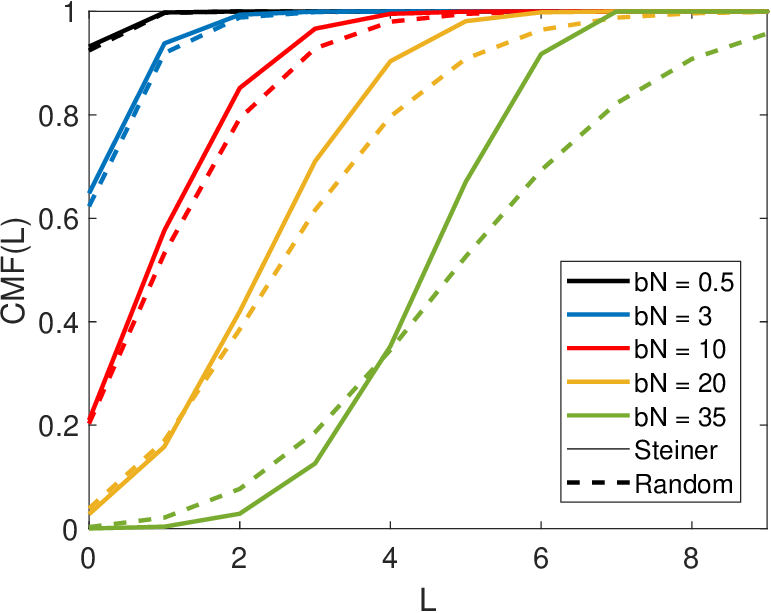}}%
\caption{Comparison of the Steiner system (solid line) and Random selection (dotted) with $K=4$ and $M=25$ in terms of their distributions of interference free slots $K'$ and number of interferers $L$. In the first two subfigures, the x-axis represents the mean traffic intensity $bN$. The third subfigure is a CMF of $L$.}
\label{fig:probabilities}
\end{figure*}

In Fig.~\ref{fig:probabilities} we compare a $S(2,4,25)$ Steiner system (solid line) and a corresponding Random selection (dashed) with the same frame length and number of repetitions.
\REV{For this Steiner system, we can have at most $C=50$ users, and at most $D=8$ collisions in a slot. In terms of the number of collision-free slots $K'$ shown in Fig.~\ref{fig:pk}, the main conclusion is that Steiner system reduces the probability of having the best ($K'=K=4$) and worst ($K'=0$ 
and $K'=1$) outcomes, while 
increasing the probability of having 'good' outcomes, such as $K'=K-1=3$, $K'=K-2=2$, for the lower values of $bN$. The ability to avoid the worst case scenarios is particularly important as the probability of $K'=0$ is tied to the performance floor. Clearly, when there are no collision-free replicas, increasing SNR is not effective and without a single packet that can be decoded, SIC cannot be applied.\footnote{The details about the reception model are provided in the next section.}}
The structure of the Steiner code becomes a disadvantage as the mean traffic intensity $bN$ increases.
However, as will become evident later, in most cases ultra-reliability cannot be achieved if the traffic intensity is too high, regardless if the access scheme is based on Steiner system or Random selection.
As such, we note that the region of interest is primarily low and medium traffic intensity, where the average number of activated users $bN<\frac{M}{2}$.

\REV{In Fig.~\ref{fig:pl} we compare the probability of encountering $L$ interferers as function of $bN$, which has an impact on the SINR in the slots where collisions occur, as well as the utility of the SIC procedure.
Having more interferers makes it less likely that all of them can be removed.
Once again, the Steiner system ensures that within the traffic intensities of interest really congested slots are rare. Note 
that with Steiner system the number of interferers is strictly limited to $D-1$, which in this case is $7$.
In~\ref{fig:cdfl}, which shows the cumulative mass function (CMF) of $L$ for different values of $bN$, one can see that unless $bN=20$ or higher, the CMF curve for the Steiner system is strictly above that of the Random selection.}

\section{Receiver Processing}\label{sec:receiver}

In this section we analyze different modes and processing techniques employed at the receiver.
The metric on which we are focusing is outage probability, as it is particularly relevant for the considered scenario.\footnote{Recall that the latency in the considered scenario is determined by the length of the frame, i.e. it is $M$ slots.} We define it as
\begin{equation}
    p_{\text{out}_i} = \mathrm{Pr} \left\{ R > \log_2\left(1+\mathrm{SINR}_{i}\right) \right\}
    \label{eq:outagep}
\end{equation}
where $R$ is the transmission rate in bits per channel use at which the packet is encoded and $\mathrm{SINR}_{i}$ denotes the final, post-processing signal-to-interference-plus-noise ratio of user $i$'s packet.
The exact definition and the means of computing it depend on the chosen scenario, which is the subject of the following subsections.
\REV{We remark that while the packets in communications requiring high reliability and low latency (e.g. URLLC) are typically short, the impact of finite blocklength effects is negligible when considering Rayleigh fading channels without channel state information at the transmitter \cite{shortpacketsurvey}. As such, we use the classical Shannon capacity limit to calculate the asymptotic outage probability.}

\subsection{Collision model}

We start with a simple model that entails a less computation-intensive processing method at the receiver.
In the collision model, collisions are assumed to be \textit{destructive}, so only the slots containing a transmission of a single device are considered.
When only user $i$ is transmitting in the slot $m$,
the received complex baseband signal in \eqref{eq:baseband} simplifies for a symbol in the slot to
\begin{equation}
    y_{m}= \sqrt{P_{x}} g_{m,i}  x_{i}+w_{m}
    \label{eq:baseband:coll}
\end{equation}
and the SNR of that signal is $\rho_{m,i}=\frac{P_{x}|g_{m,i}|^2}{\sigma^2} = \theta |g_{m,i}|^2$.
Since channel coefficients are Rayleigh distributed random variables (r.v.'s), the SNR of each packet follows exponential distribution $f_\text{exp}(\rho;\theta)= \frac{1}{\theta}e^{-\frac{\rho}{\theta}}$.
As each device transmits $K$ times, there might be up to $K' \leq K$ collision-free replicas, the probability of which is given by \eqref{eq:collfree:random} and \eqref{eq:collfree:steiner}, and it is possible to combine the signals to improve the overall SNR.
Denote by $\mathcal{J}_i$ the set of indices corresponding to the uninterfered transmissions of device $i$.
Using MRC, we obtain 
\begin{equation}
    \sum_{j\in\mathcal{J}_i} g_{j,i}^* y_j=\sqrt{P_x} x_i \sum_{j\in\mathcal{J}_i} |g_{j,i}|^2 + \sum_{j\in\mathcal{J}_i} g_{j,i}^* w_j
\end{equation}
that yields the SNR $\rho_{i}= \theta \sum_{j\in\mathcal{J}_i} |g_{j,i}|^2$. As a sum of exponentially distributed r.v.'s, the total SNR after combining, conditioned on $K'$, has a gamma distribution $f_\text{gam}(\rho;K',\theta)=\frac{1}{\Gamma(K') \theta^{K'}} \rho^{K'-1} e^{-\frac{\rho}{\theta}}$.
As follows from \eqref{eq:outagep}, the decoding is unsuccessful whenever $\rho_i < 2^R-1$, hence
\begin{equation}
p_\text{out}(R, \theta,U)\!= p_{\mathrm{CF}}(0|U) +\! \sum_{K'=1}^{K}\!\!F_\mathrm{gam}(2^R-1;K',\theta)p_{\mathrm{CF}}(K'|U)
\label{eq:outp:coll}
\end{equation}
where $p_{\text{CF}}(\cdot|U)$ is given by either \eqref{eq:collfree:random} or \eqref{eq:collfree:steiner}, depending on whether random or Steiner patterns are used, respectively. 
The equation can be further marginalized over $U$ to account for the specific activation process.

\begin{figure}[t!]	
	\centering
	\includegraphics[width=0.85\linewidth]{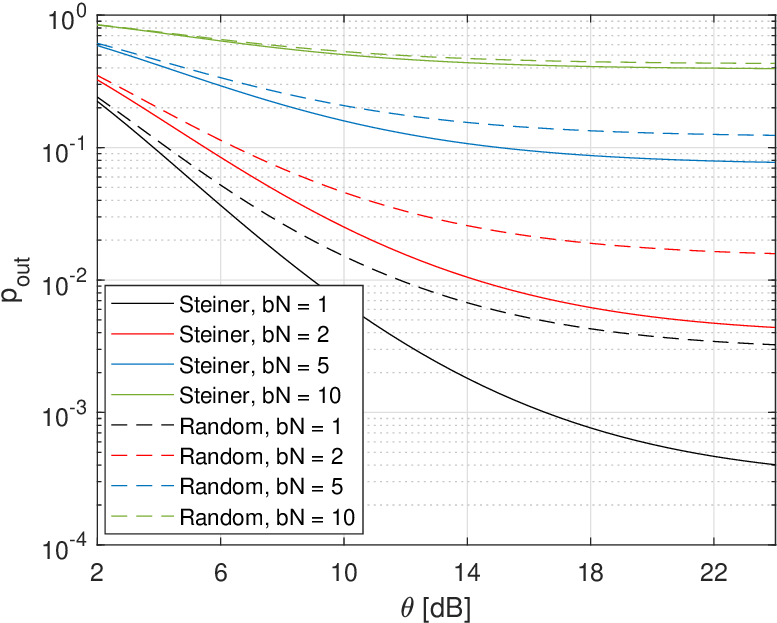}
	\caption{Outage probability performance of the system employing random and deterministic patterns as a function of the average received SNR for different mean number of active devices $bN$.}
	\label{fig:collision_single_steiner_v_random}
\end{figure}

In Fig.~\ref{fig:collision_single_steiner_v_random} we present the results based on \eqref{eq:outp:coll} that show the performance of deterministic patterns based on Steiner system, and random selection. The parameters chosen are: $M=25$ slots, $K=4$ repetitions and $N=C=50$ users and Steiner patterns.
The activation probabilities are $b=[0.02, 0.04, 0.1, 0.2]$ that translate to the mean number of active devices in a frame equal to $[1, 2, 5, 10]$ respectively. The transmission rate is $R=2 \frac{bit}{c.u.}$. 
The properties of the Steiner system 
discussed in the previous section lead to tangible gains in terms of outage probability; up to an order of magnitude 
when compared to Random selection. 
In the next section, we employ more sophisticated processing that further decreases the outage probability.

\subsection{Collision model with successive interference cancellation}

A natural
method to improve receiver performance is to apply 
SIC. We thus remove packets which were successfully decoded from the received signal, including all of their $K$ replicas. This has the potential to greatly improve the performance, as it allows to remove the interference from the slots where collisions occurred and make them usable in the subsequent iterations of the decoding process.

A rigorous analysis of the models involving SIC is known to be inherently difficult\cite{csa1} and the exact analytical results typically do not exist except for asymptotic cases and some simple special cases.
Due to the multitude of possible configurations of the selected transmit patterns and different dependencies they create, the problem became intractable already when $U>4$.
This motivated us to look for approximate results.

For this, we consider the first few rounds of  SIC. In each round, we condition on the number of decoded users in the preceding rounds.
Due to the unique structure of the Steiner system, which ensures thatthe access patterns of active users
cover the frame uniformly, it is possible to simplify some of the steps by using averages.
Indeed, rather than having to sum/integrate over all possible outcomes of many variables, we can work with the averages in terms of the number of collision-free slots, the combined SNR, etc. 
This is in contrast to Random selection, where the covering is uniform with high probability only when there are many active users, 
while for low $U$ the patterns can have significant overlap.
Consequently, our approach is suitable only for Steiner systems and will not provide a good approximation if the patterns follow a Random selection.


As above, the analysis will be performed from a point of view of an arbitrary user. 
First, consider the case that $l_1=1, \dots, U-1-S$ users are decoded in the first iteration of SIC.
Here, $S$ denotes the number of users who for some reason are excluded from the procedure and cannot be cancelled (this will be explained in detail later).
For convenience we will introduce auxiliary variable $u_S=U-1-S$.
If we treat all the $u_S$ users independently, each with a probability of outage $p_\text{out}(R, \theta,U)$ given by \eqref{eq:outp:coll}, the distribution of $l_1$ can be approximated\footnote{Note that in general the decoding events are not independent. Eq.~\eqref{eq:outp:coll} is a weighted mean of all realizations of $K'$, however it is not possible to have a situation with 2 active users where $K'_1\neq K'_2$.} 
with binomial distribution $f_{\text{bin}}(l_1;u_S,1-p_\text{out}(R,\theta,U))$.
For the remaining $u_S-l_1$ users that were not successful in the first round of SIC, it is important to account for the fact that they nevertheless accumulated some (but less than $2^R-1$) SNR already. 
Given $K'$, their SNRs are drawn from a truncated gamma distribution $\frac{f_\text{gam}(\rho;K',\theta)}{F_\text{gam}(2^R-1;K',\theta)}$.
By taking its mean and marginalizing over $K'$, we can determine the mean residual SNR, i.e. the amount of signal power that is missing before the packet can be decoded:
\begin{equation}
\rho_\text{res} = \sum_{K'=0}^{K}\left( 2^R-1 - \theta \frac{ \gamma\left( \frac{2^R-1}{\theta},K'+1\right)}{\gamma\left( \frac{2^R-1}{\theta},K'\right)} \right) p_{\text{CF}}(K'|U)
\label{eq:residSNR}    
\end{equation}
where $\gamma(s,x)=\int_0^x t^{s-1}e^{-t}dt$ is a lower incomplete gamma function.

The expected number of collision-free slots per user
is $\widehat{K}(U)=\sum_{K'=0}^{K}K'p_{\text{CF}}(K'|U)$. 
Since all the collision-free slots from the first round have already been taken into account, in the second round we are only interested in new slots that become collision-free after cancelling the $l_1$ successful users of the first round.
On average, there will be $\widehat{K}(U) - \widehat{K}(U-l_1)$ new slots, so the probability of decoding a packet in the second round of SIC is
\begin{equation}
  p_{\text{out},2}(l_1) =F_\text{gam}\left(\rho_\text{res};\widehat{K}(U) -\widehat{K}(U-l_1),\theta\right)
    \label{eq:outp2}
\end{equation}
Similarly, we then consider the number of additional messages that can be decoded in the second iteration $l_2=0,1,...u_S-l_1$ which is given by $f_\text{bin}\left(l_2;u_S-l_1,1-p_{\text{out},2}(l_1)\right)$.
We halt this procedure at the third iteration.
At this point the reference user observes the system with $U-l_1-l_2$ devices (including itself), however, since there was no attempt to decode its packet yet, it is subject to $p_\text{out}(R, \theta,U-l_1-l_2)$. 
By marginalizing over $l_1$ and $l_2$, the outage probability conditioned on $S$ is then
\begin{equation}
\begin{aligned}
    &p_{\text{out},\text{SIC}}(R,U|S) \\
    &\;\; =  p_\text{out}(R,\theta,U)^{U-S} + \sum_{l_1=1}^{u_S} f_\text{bin}\bigl(l_1;u_S,1-p_\text{out}(R,\theta,U)\bigr) \\ 
    &\;\;\;\;\; \times \!\!\! \sum_{l_2=0}^{u_S-l_1} \!\!f_\text{bin}\bigl(l_2;u_S-l_1,1\!-p_{\text{out},2}(l_1)\bigr)\, p_\text{out}(R,\theta,U\!-l_1\!-l_2)
\end{aligned}
\label{eq:outp:sic_part}
\end{equation}
where the first term corresponds to the case in which all users fail and SIC cannot proceed.

The expression \eqref{eq:outp:sic_part} with $S=0$ approximates very well the simulation results when the number of active devices $U$ is low.
As $U$ grows, the approximation and simulations start to diverge in the high SNR regime, with the latter exhibiting plateauing.
The reason is due to the existence of \textit{stopping sets} \cite{stopping2002}.
It is easy to imagine a situation where the access patterns overlap in such a way that there are no collision-free slots and consequently SIC cannot be applied.
Formally, a stopping set $s^{(n)}$ of order $n$ is a subset of $n$ patterns such that in every slot there is either $0$, or $\geq 2$ users; that is, the decoding cannot proceed as there is no slot with a single transmission only.
Further, denote by $T^{(n)}$ the set of all stopping sets of order $n$ for a given Steiner system and by $|T^{(n)}|$ its cardinality\footnote{While it would be more precise to  write $T_{S(t,K,M)}^{(n)}$, for brevity we decide to drop the subscript $_{S(t,K,M)}$ (as we similarly do in the case of quantities $C$ and $D$). In this paper we always consider a single Steiner system at a time so this should not lead to any confusion.}. 
In order to take into account stopping sets and augment the expression \eqref{eq:outp:sic_part}, we need to consider three cases. 
If there is a stopping set of certain order $n$, then with probability $n/U$ the user in focus is its member and cannot be decoded. 
Conversely, with probability $1-n/U$ the user is not involved in that stopping set and decoding is possible, however SIC is impaired since there are $S=n$ noncancellable users.
Otherwise, if there are no stopping sets then $S=0$ and there are no limitations on SIC. Combining those three cases and \eqref{eq:outp:sic_part}, we obtain the following approximation for the outage probability with SIC:
\begin{equation}
\begin{aligned}
    p_{\text{out},\text{SIC}}(R,U) = & \sum_{n \in \mathfrak{N}} q_{1}(n|U) \!\! \left( \frac{n}{U} + \frac{U-n}{U} p_{\text{out},\text{SIC}}(R,U|n) \right) \\
    & + p_{\text{out},\text{SIC}}(R,U|0) \left(1- \sum_{n \in \mathfrak{N}} q_{1}(n|U) \right)
    \label{eq:outp:sic}
\end{aligned}
\end{equation}
where summation is over $\mathfrak{N}=\{n:T^{(n)} \neq \emptyset \}$ and $q_{1}(n|U) \approx f_{\text{bin}} \big( 1;\binom{U}{n},|T^{(n)}| / \binom{C}{n} \big)$ is the probability that there is a stopping set of order $n$ among $U$ active users. We note that this is an approximation, because the $\binom{U}{n}$ tuples are not independent. 
It is important to understand 
that we are interested in 
{\it exactly one}
stopping set of a given order, not 
``at least one''. The reason is that stopping sets are closed under union, so their combination produces another stopping set of a higher order \cite{stopping2002}. As such, by summing over $n$ we would count some of the stopping sets multiple times.

Furthermore, in \eqref{eq:outp:sic} it is not necessary to sum over whole $\mathfrak{N}$ to obtain a good approximation. In practice, 
performance is impacted primarily by the stopping sets of the lowest existing order, which we denote by $n'$. They are decisive for two reasons. 
For a given Steiner system, the outage probability cannot be made arbitrarily low regardless of the SNR whenever $U\geq n'$.
Secondly, simulations show that for $i<j$, $q_{1}(i|U) > q_{1}(j|U) $ and the difference can reach several orders of magnitude, making $q_{1}(n'|U)$ dominant overall.
This is fortunate, since finding $T^{(n)}$ requires an exhaustive search, which for high $n$ becomes prohibitive.
Consequently, when generating results, for each Steiner system we use only the stopping sets of the lowest existing order $n'$, and $n'+1$ whenever $T^{(n'+1)}$ is not empty. If $T^{(n'+1)} = \emptyset$, $T^{(n')}$ is sufficient.

\begin{figure}[t!]	
	\centering
	\includegraphics[width=0.85\linewidth]{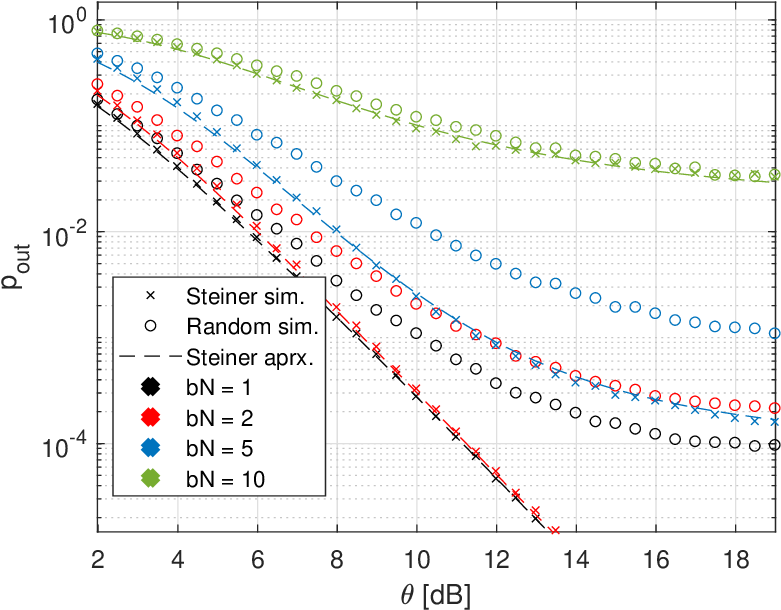}
	\caption{Comparison of the simulation results involving the exact procedure and the proposed approximation}
	\label{fig:collision_sic_steiner_approx}
\end{figure}

In Fig.~\ref{fig:collision_sic_steiner_approx} we plot the outage probability as given by the proposed approximation \eqref{eq:outp:sic} (dashed lines) and compare it to the results of the corresponding simulations in which the full procedure is implemented (markers). The derived approximations prove to be very close to the exact results across the whole SNR range and different traffic intensities.
The improvement compared to a system without SIC (cf. Fig.~\ref{fig:collision_single_steiner_v_random}) is significant and allows to achieve ultra reliability at much lower SNRs. Particularly important is the fact that Random selection exhibits a performance floor even when the mean traffic intensity is as low as $1$ user/frame. This is a consequence of the stopping sets, which in a Random selection can occur already with two users if they select exactly the same pattern. This has been discussed also in \cite{Boyd2018}. 
Conversely, in a Steiner system the number of collisions between two patterns is strictly limited, hence there are no stopping sets as long as the number of active users is sufficiently small.
Specifically, it is easy to show that with a maximum of $t-1$ collisions and $K$ repetitions at least 
\begin{equation}
    n' \geq \left\lceil \frac{K}{t-1} \right\rceil+1
\end{equation}
users need to be active for the stopping set to occur. 
In practice, for some Steiner systems that number is even higher, e.g. in the used $S(2,4,25)$ no subset of size $<7$ exist that would form a stopping set. 
Those issues, as well as the results for other systems, are further discussed in Section~\ref{sec:sys_design:resources}. 

\subsection{Model with Full MRC}

In the following we will consider a more involved model, in which the receiver is capable of using the totality of all the replicas (including those experiencing interference) and combines them using MRC.

Without loss of generality, let us consider an arbitrary active user $i$ and one of its transmissions $j \in \{ m: V_{m,i} = 1\}$. The SINR of this signal

\begin{equation}
    \rho_{j,i} = \frac{P_x |g_{j,i}|^2}{\sum_{k\in {\cal U}\setminus \{i\}} V_{j,k} P_x |g_{j,k}|^2 + \sigma^2}
\end{equation}
is an r.v. Here $\cal U$ is the set of all active users. 
Assuming there are $L$ interferers in slot $j$, i.e. 
$\sum_{k\in {\cal U}\setminus \{i\}} V_{j,k} = L$,
we can denote this SINR as $\frac{X}{Y+1}$, where $X$ follows the exponential distribution $f_\text{exp}(x;\theta)$ and $Y$ the gamma distribution $f_\text{gam}(y;L,\theta)$. Hence, 
\begin{equation}
\begin{split}
    &P\left(\frac{X}{Y+1}>z\right) \\
    &\quad = \int_{0}^{\infty} \left( \int_{(y+1)z}^{\infty}  \frac{1}{\theta} e^{-\frac{x}{\theta}} dx \right) \frac{1}{\Gamma(L)\theta^{L}} y^{L-1} e^{-\frac{y}{\theta}} dy \\
    &\quad = \frac{1}{\Gamma(L)\theta^L} \int_{0}^{\infty} y^{L-1} e^{-\frac{y}{\theta}} \cdot \left. -e^{-\frac{x}{\theta}} \right|_{(y+1)z}^{\infty} dy \\
    &\quad = \frac{e^{-\frac{z}{\theta}}}{\Gamma(L)\theta^L} \int_{0}^{\infty} y^{L-1} e^{-\frac{y(z+1)}{\theta}} dy \\
    &\quad = \frac{e^{-\frac{z}{\theta}}}{\Gamma(L)\theta^L} \cdot \frac{(L-1)! \theta^L}{(z+1)^L} = \frac{e^{-\frac{z}{\theta}}}{(z+1)^L}
\end{split}
\end{equation}
where the last integral can be computed by integrating it by parts $L-1$ times. Finally, by taking the derivative of $1-P\left(\frac{X}{Y+1}>z\right)$ we obtain the pdf of the SINR given $L$ interferers:
\begin{equation}
    f_{SINR}(z;\theta | L) = \frac{e^{-\frac{z}{\theta}}(\theta L + z + 1)}{\theta(z+1)^{L+1}}
\end{equation}
and it can be seen that for a special case of $L=0$ the expression reduces to a simple exponential distribution.

\REV{Because $L$ is a random variable, it needs to be marginalized out.
To do that, let us condition on the number of interference-free slots $K'$ (given by \eqref{eq:collfree:random}, \eqref{eq:collfree:steiner}).
Notice, that fixing $K'$ has an implications for the distribution of interferers in the remaining slots.
Namely, by fixing $K'$ we implicitly exclude some patterns (in case of Steiner system) or slots (in case of random selection). Hence, we introduce the modified version of the expressions \eqref{eq:interferers:random}, \eqref{eq:interferers:steiner}:}
\begin{equation}
	p_{\mathrm{I,R}}(L|K',U) = \frac{ \binom{M-1-K'}{K}^{U-1-L} \binom{M-1-K'}{K-1}^{L} \binom{U-1}{L}}{\binom{M-K'}{K}^{U-1} }
	\label{eq:2}
\end{equation}
\begin{equation}
	p_{\mathrm{I,S}}(L|K',U) =  \frac{\binom{D-1}{L} \binom{C-1 -(D-1)(K'+1)}{U-1-L}}{\binom{C-1-(D-1)K'}{U-1} }~.  \label{eq:4}
\end{equation}

Secondly, if the slot contains interference, then by definition $L>0$ so the case $L=0$ has to be excluded and the distribution re-normalized. Taking all this into account, the distribution of the SINR in the interfered slot becomes

\begin{equation}
    f_{I-SINR}\big(z;\theta|U,K'\big)=\sum_{L=1}^{U-1}f_{SINR}(z;\theta|L)\frac{p_\mathrm{I}(L|K',U)}{1-p_\mathrm{I}(0|K',U)}~.
    \label{eq:i_sinr}
\end{equation}
In order to obtain the distribution of the total SINR, we combine the above with the contribution from $K'$ interference-free slots and marginalize out $K'$:
\begin{equation}
\begin{aligned}
    f_{tot,SINR}\big(z;\theta|U\big) &= \!\! \sum_{K'=0}^{K}p_{\mathrm{CF}}(K'|U)f_\mathrm{gam}(z;K',\theta) \\
    &\;\;\; \ast  f_{I-SINR}(z;\theta|U,K')^{\ast (K-K')}
    \label{eq:tot_sinr}
\end{aligned}
\end{equation}
\REV{where $\ast$ denotes convolution}, $f^{\ast n} \stackrel{\text{def}}{=} \underbrace{f\ast \dots \ast f}_{n}$ and $f^{\ast 0}=\delta$, with $\delta$ denoting the Dirac delta distribution. Similarly, we set $f_{\text{gam}}(z;0,\theta)=\delta$ to keep the above expression \eqref{eq:tot_sinr} compact. Finally, we have that
\begin{equation}
\begin{aligned}
    p_{out,cap}(R,\theta,U) &= \!\! \sum_{K'=0}^{K}p_{\mathrm{CF}}(K'|U) F_\text{gam}(z;K',\theta) \\
    &\;\;\; \ast \left.  f_{I-SINR}(z;\theta|U,K')^{\ast (K-K')} \right|^{z=2^R-1}
    \label{eq:outp_capture_single}
\end{aligned}
\end{equation}

While not excessively complex, evaluation of \eqref{eq:outp_capture_single} requires $(K-1)(K+2)/2$ numerical integrations. This is not an issue for the values of $K$ considered in this work; however, to address this we provide in Appendix \ref{sec:appendix:} an even simpler approximation that avoids the convolutions altogether.

\begin{figure*}[t!]%
\centering
\subfigure[]{
\label{fig:capture_single}%
\includegraphics[width=0.45\linewidth]{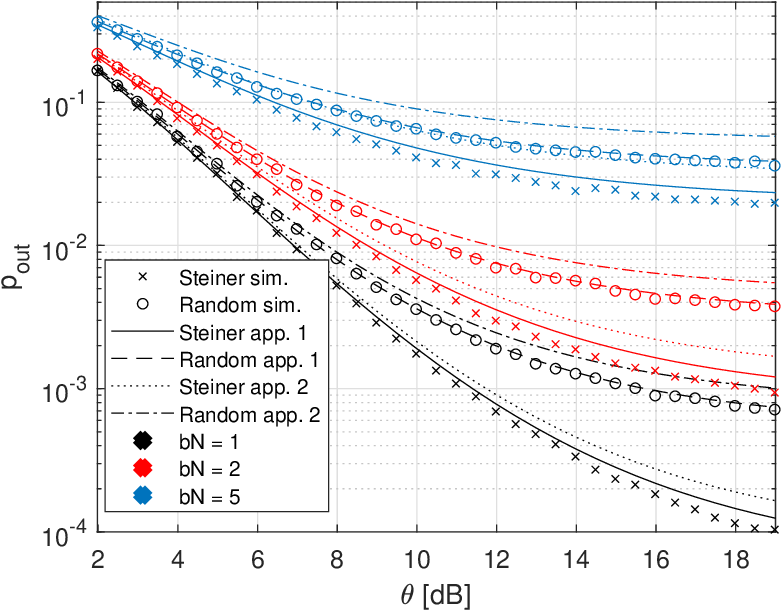}}
\hfill
\subfigure[]{
\label{fig:capture_sic}%
\includegraphics[width=0.45\linewidth]{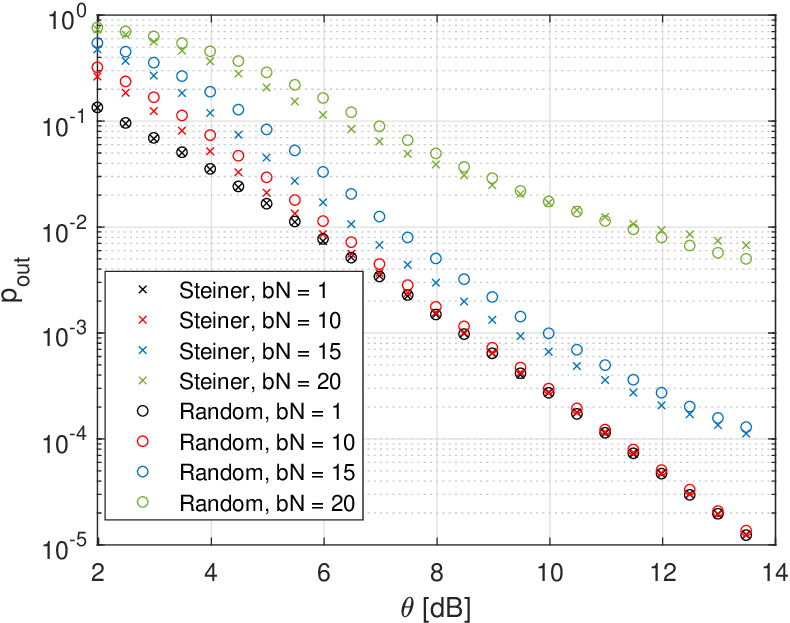}}
\hfill
\caption{Outage probability in the Full MRC model (a) without SIC and (b) with SIC. }
\label{fig:capture}
\end{figure*}

In Fig.~\ref{fig:capture_single} we present a comparison of the outage probability obtained by the simulation results and the developed approximations.
Once again, we consider the performance of the Steiner system and Random selection across the range of received SNRs and traffic intensities.
The results from simulations, which implement the exact procedure, are given with markers.
Solid and dashed lines (Steiner and Random respectively) correspond to the approximation based on eq.~\eqref{eq:outp_capture_single} (Approx. 1). Similarly, dotted and dash-dotted lines (Approx. 2) correspond to the simpler approximation by gamma distribution discussed in the Appendix \ref{sec:appendix:}. 
Approx. 1 follows very closely the actual simulation results, however we note that in case of Steiner system the deviation is slightly larger than for Random selection.
Albeit being simpler, the accuracy of Approx. 2 is not far off,
especially for lower traffic intensities $bN$, which are of primary interest when high reliability is considered.

In Fig.~\ref{fig:capture_sic} we show the simulations results for the Full MRC model that leverages SIC.
The improvement over non-SIC, cf. Fig.~\ref{fig:capture_single}, processing is significant as the performance does not exhibit plateauing and is capable of achieving ultra-reliability even for traffic intensities as high as $bN=10$.
The superiority of Steiner system over Random selection, especially for high SNRs, is diminished; this aspect is further discussed in Section~\ref{sec:sys_design:random}.

\section{System Design favors Steiner Sequences over Random Selection}\label{sec:sys_design:random}

Until now, the implicit assumption was that perfect channel estimates have been available. In reality channel estimates are never perfect, however, the quality of estimation can be improved by making the pilot sequences longer and/or investing in them more transmit power, as long as the pilot sequences of transmitting users are kept orthogonal. 
Collisions in the pilot domain are particularly problematic as they lead to pilot contamination and, consequently, very poor channel estimates. Typically, that means proper equalization is not possible and the packet replicas involved in the collision are unusable, i.e. cannot be combined with others through MRC or removed with SIC\footnote{Note that even if the packet can be decoded based on replicas from other slots, it cannot be removed from the slots where pilot collisions occurred as the channel estimates are not available.}.
This is a significant challenge for random access schemes that rely on fully random selection of access patterns.
Since any user can be active in any slot, the only way to avoid pilot collisions, would be to assign a unique orthogonal sequence to each of the $N$ devices.
In many cases, however, this is not feasible or practical (e.g. with large population of intermittently active devices similar to the scenario addressed in this work).
Instead, a common approach is to provide a pool of $Q<N$ pilot sequences from which users pick one at random every time they become active and accept, that some collisions will inevitably occur.

In that case, it is possible to provide a reasonable lower bound for the model utilizing SIC as $\theta \rightarrow \infty$.
Let us again focus on an arbitrary user $u$ and one of its slots. We can distinguish two types of events.
In the first case, whenever one or more interferers select the same pilot sequence as the user of interest $u$, the packet replica is lost. This is given by
\begin{equation}
    p_{(\text{I})} = \sum_{L=1}^{U-1} p_\mathrm{I}(L|U) \left(1-\left(1-Q^{-1} \right)^L\right)~.
\end{equation}
The second type of event, is when user $u$'s pilots are intact, however there are some pilot collisions among the interferers.
The probability that the slot is of this second kind is
\begin{equation}
    p_{(\text{II})} = \! \sum_{L=2}^{U-1} p_\mathrm{I}(L|U) \bigg( \! 1 - \frac{\prod_{i=0}^{L-1} Q -i}{Q^L} - \Big(1 -\left(1-Q^{-1} \right)^L \Big) \! \bigg)~.
\end{equation}
In that case, due to the lack of channel knowledge SIC cannot be applied to remove the interferers,
so the SINR is limited to at most $\frac{X}{Y_1 +\dots+ Y_{L'} +1}$, where $L'$ is the number of mutually colliding (in the pilot domain) interferers. 
For the lower bound, we can fix $L'=2$ and, as $\theta \rightarrow \infty$, the '$1$' in the denominator can be dropped.
Along with the fact that $X$, $Y_1$, and $Y_2$ are exponentially distributed with the same scale parameter $\theta$, the SINR of user $u$'s replica follows a beta prime distribution $f_{BP}(x; \alpha,\beta)=\frac{x^{\alpha-1}(1+x)^{-\alpha-\beta}}{B(\alpha,\beta)}$ with $\alpha=1$ and $\beta=2$. 
Considering that there can be $n=0,\dots, K$ such replicas (with the remaining ones being of the first type - i.e. lost due to pilot collisions), the outage probability can be bounded by
\begin{equation}
    p_{bound}(R,U) = p_{(\text{I})}^K + \sum_{n=1}^{K} \! \binom{K}{n} p_{(\text{I})}^{K-n} p_{(\text{II})}^{n} F_{BP}(2^R-1;n,2)
    \label{eq:out_bound}
\end{equation}
where $F_{BP}(\cdot;\cdot,\cdot)$ is the CDF of beta prime distribution and we leverage the fact that sum of its $n$ i.i.d variables is also beta-prime distributed, in this case, with $\alpha_n=n$, $\beta_n = 2$. 

Unlike Random selection, the Steiner system guarantees that at most $D$ users can be active in any given slot. Furthermore, since each user has to be assigned a specific access pattern for their packet replicas, it can be simultaneously instructed which pilot sequence to use in which slot, thus eliminating any possibility of collisions with just $Q=D$ orthogonal pilot sequences. Recalling that in a Steiner system the number 
supportable users is $N=C=\binom{M}{t}/ \binom{K}{t}$, while $D=\binom{M-1}{t-1} / \binom{K-1}{t-1}$, we have that $D = N\frac{K}{M}$, i.e. the number of pilot sequences required to ensure no collisions is reduced by a factor $K/M$ compared to the Random selection.

Another caveat is that, clearly, the receiver must know where each replica of each user is located in order to perform combining through MRC. Because with the Steiner system there is an association between pilot sequences and user IDs, observing a certain pilot sequence in a given slot automatically indicates which user is active and where to look for its remaining replicas. This is not the case when Random selection scheme is used so additional procedures might be required. One possibility is to look for the correlation between signals in different slots and combine those with the highest correlation score.
However, such a solution is not perfect as it might miss some of the replicas or introduce false positives. Furthermore, it entails exhaustive search and, hence, high complexity.
Alternatively, a unique ID that can be decoded independently of the rest of the payload could be added to each packet or, each slot could be preceded by an activity-indication phase. Clearly, the downside of this solution is the introduction of overhead.

\begin{figure}[t!]	
	\centering
	\includegraphics[width=0.85\linewidth]{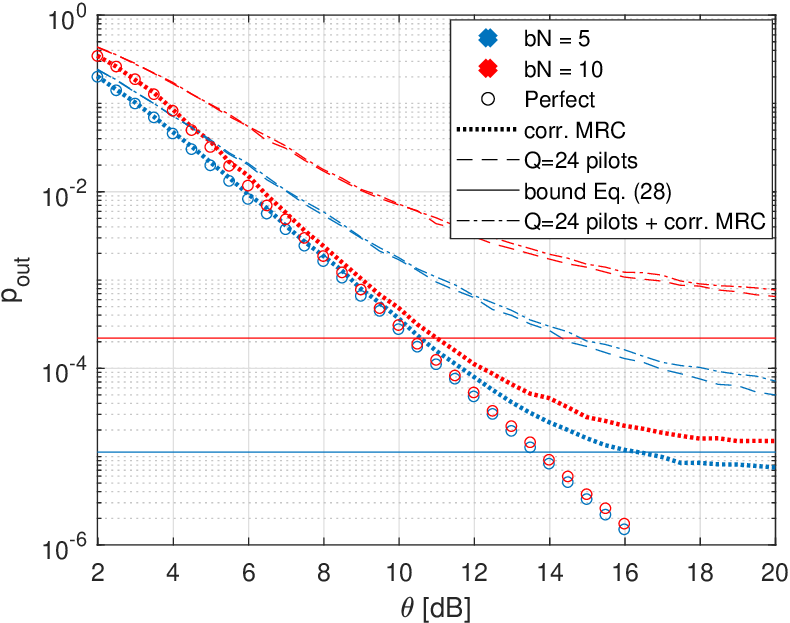}
	\caption{Impact of pilot collisions and realistic MRC processing on Random selection scheme. Although not shown here for the sake of readability, the performance of Steiner system matches, and for low $\theta$ even exceeds, the perfect Random scheme.}
	\label{fig:imperfect_random}
\end{figure}

Lastly, we remark that MRC processing is impeded when using Random selection.
Note that the SINR of the combined packet is equal to the sum of the SINRs of its constituents only when the interference in each is uncorrelated. This will not be the case if a given user collides with another in more than one slot. Whether the resulting SINR will be higher or lower than the sum will depend on whether the interference adds destructively or constructively.
Interestingly, even though both are equally likely, the performance of SIC is ultimately impaired\footnote{Consider the case with two users colliding in more than one slot and let us establish a baseline where the combined SINR is equal to the sum of SINRs in individual slots. As already mentioned, in reality, the SINR after MRC will be lower (we can call it negative MRC) or higher (positive MRC) than the baseline. Importantly, if the combining is negative (positive), it is negative (positive) for both users. Consider now 4 possible decoding outcomes than can happen in the baseline scenario. If both users are successful, then positive MRC would have no effect. Similarly, if only one of them succeeds but the other fails, positive MRC also wouldn't change anything since the successful user can be cancelled through SIC anyway. Only when both users fail the positive MRC can make a difference - that is if it makes at least one of them decodable and triggers SIC.
Now let us consider negative MRC. When both users fail it has no effect. If there is one successful user, it can happen that negative MRC turns it into an undecodable one, thus making SIC impossible. Similarly, if both users are successful, negative MRC could make them both undecodable (in this case it is not enough to turn just one of them, as the SIC could still be applied). 
In the end, even though negative and positive MRC are equally probable, if the system uses SIC, the negative MRC has the potential to be detrimental in three out of four cases, while the latter can only help in one of them.}. To circumvent that, a more computationally-heavy equalization method such as zero-forcing (ZF) or minimum mean squared error (MMSE) would have to be used.
Conversely, in a Steiner system with $t=2$ each user is guaranteed to collide with another at most once so the interference is always uncorrelated.

In Fig. \ref{fig:imperfect_random} we demonstrate the impact of the above described issues.
The markers represent the performance of the probability of the outage for the idealized version of the Random selection scheme and serve as a reference.
The dotted lines show the actual performance of MRC in the presence of correlated interference. The dashed lines depict the scenario with finite pool of $Q=24$  pilot sequences and the solid horizontal line is the corresponding lower bound as given by \eqref{eq:out_bound}. Lastly, dash-dotted curves take into account both the corralated interference and the correlated MRC. 
The Fig.~\ref{fig:imperfect_random} reveals that, in a more realistic setting, the performance of the Random selection would be significantly impaired and cannot match that of Steiner system, cf. Fig.~\ref{fig:capture_sic}.  

\section{Performance Evaluation: Choice of Frame Parameters $M$ and $K$}\label{sec:sys_design:resources}
\begin{table}[t!]
 \renewcommand{\arraystretch}{1.2}
 \centering 
 \caption{Properties of Steiner Systems}
 \begin{tabular}{| c | c | c | c |}
    \hline
     & $C$ & $D$ & $T^{(n)}$\\ \hline
    $S(2,5,25)$ & $30$ & $6$ & $|T^{(9)}| = 1150$ \\ \hline
    $S(2,5,41)$ & $82$ & $10$ & $|T^{(6)}| = 41$,  $|T^{(7)}| = 0$ \\ \hline
    $S(2,4,25)$ & $50$ & $8$ & $|T^{(7)}| = 266$, $|T^{(8)}| = 1827$ \\ \hline
    $S(2,4,37)$ & $111$ & $12$ & $|T^{(6)}| = 37$,  $|T^{(7)}| = 0$ \\ \hline
    $S(2,3,25)$ & $100$ & $12$ & $|T^{(4)}| = 4$, $|T^{(5)}| = 92$ \\ \hline
    $S(2,3,33)$ & $176$ & $16$ & $|T^{(4)}| = 429$, $|T^{(5)}| = 77$ \\ \hline
  \end{tabular}
  \label{tab:params}
\end{table}

Lastly, we consider Steiner systems with different configurations of the frame length $M$ and number of repetitions $K$. In Table~\ref{tab:params} we provide the relevant parameters for the systems used in this work, namely the number of patterns $C$, maximum number of interferers per slot $D$, order of the smallest existing stopping sets as well as their number. The patterns themselves 
can be found in \cite{LJCR}.
The objective is to determine the highest supported rate $R$ for a given traffic intensity $bN$ and fixed mean SNR $\theta$, that fulfills certain target reliability $\epsilon_{tar}$:
\begin{maxi}|s|[2]           	
    {}           	
    {  \mathrlap{R}\phantom{aaaaaaaaaaaaaaaaaaaaaaaaaa}
    }   
    {\label{eq:opt2}}   	
    {}                                
    \addConstraint{ \sum_{u=0}^{N} f_{\text{bin}}(u,b,N) p_{\text{out}}(R,\theta,u)} {\leq \epsilon_{tar}{}
    }    
\end{maxi}

In addition, we define the spectral efficiency of the system given by $bN \cdot R/M$. 
In the figures, the results are plotted as a function of the absolute mean traffic intensity $bN$. We note that since $N$ is different for each Steiner system, so is the activation probability $b$.
Furthermore, in order to jointly compare Steiner systems with different number of repetitions $K$, we decide to normalize their mean received SNRs. 
The rationale is that, with $\theta$ being the same in each case, the systems with higher $K$ would use proportionally more energy and thus have an advantage.
Consequently, we perform our evaluations by fixing $\theta K = 25$ dB.
Finally, we set the target for the outage probability to $\epsilon_{tar}=10^{-5}$.

We focus on two cases a) the Full MRC model without SIC and b) the collision model with SIC.
The results are obtained based on the derived approximations \eqref{eq:outp_capture_single} and \eqref{eq:outp:sic}, respectively, which are applied to \eqref{eq:opt2}. 
To improve the readability of the figures, we do not show the results for Random selection schemes in this section, noting that they are always strictly worse than the corresponding Steiner system (see earlier discussions and Figs.~\ref{fig:collision_single_steiner_v_random}-\ref{fig:capture}).

\begin{figure}[t!]	
	\centering
	\includegraphics[width=0.85\linewidth]{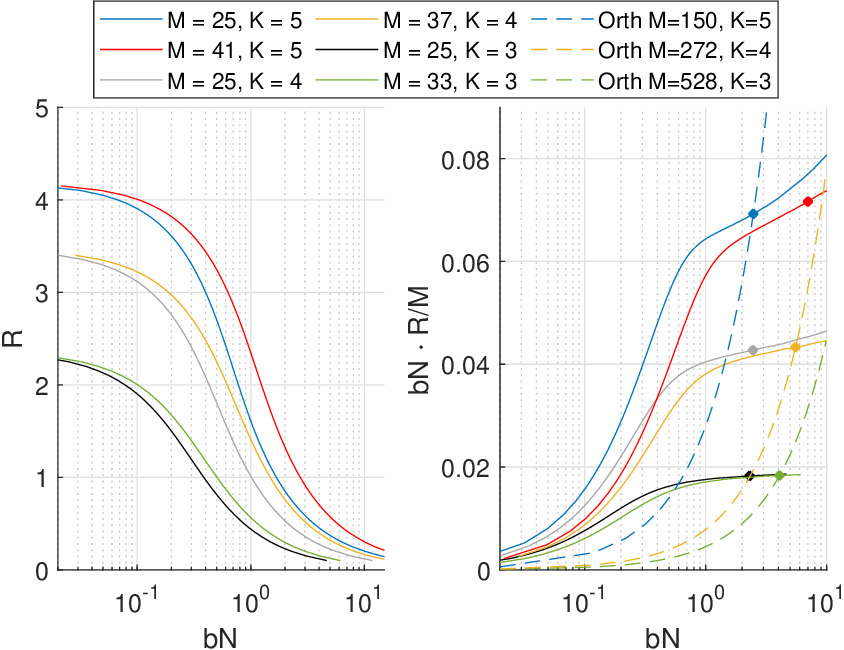}
	\caption{ Full MRC model without SIC. (a) Maximum rate for which outage probability target $p_{out} \leq \epsilon_{tar} = 10^{-5}$ is fulfilled as a function of the traffic intensity, and (b) corresponding spectral efficiency.}
	\label{fig:bn_combined_cap}
\end{figure}

We start with the rate $R$ of the Full MRC model shown in Fig.~\ref{fig:bn_combined_cap}(a). As expected, for a given mean number of active users, the larger the frame and the number of repetitions, the higher the rate. Increasing the frame size decreases the chance of collisions, while increasing $K$ makes the transmission more robust and allows to harvest more diversity. 
The relationship, however, is not as straightforward when it comes to spectral efficiency shown in Fig.~\ref{fig:bn_combined_cap}(b). For a given number of repetitions $K$, increasing the frame length actually reduces the spectral efficiency, as the increase in rate is not enough to offset the extra resources 
( $S(2,5,25)$ vs $S(2,5,41)$ or $S(2,4,25)$ vs $S(2,4,37)$). 
Furthermore, even though higher $K$ itself is generally beneficial, the system with high $M$ and $K$ might be less spectrally efficient than the one with lower parameters when $bN$ is low 
($S(2,5,41)$ vs $S(2,4,25)$).

\begin{figure}[t!]	
	\centering
	\includegraphics[width=0.85\linewidth]{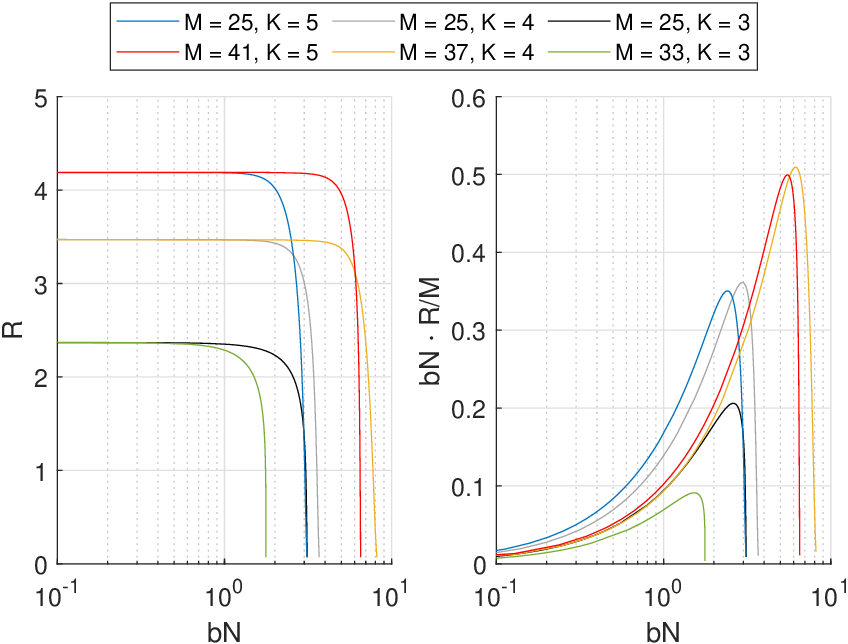}
	\caption{Collision model with SIC. (a) Maximum rate for which outage probability target $p_{out} \leq 10^{-5}$ is fulfilled as a function of the traffic intensity, and (b) corresponding spectral efficiency.}
	\label{fig:bn_combined_col}
\end{figure}

To provide further insight, in Fig.~\ref{fig:bn_combined_cap}(b) we also mark the traffic intensity $bN$ beyond which orthogonal resource allocation becomes more spectrally efficient than Steiner system. 
To find this value, first, we note that when resources are orthogonal, the maximum rate does not depend on the traffic intensity and is given by $R_{orth} = \log_2 (F^{-1}_{gam}(\epsilon_{tar};K,\theta) + 1)$, where $F^{-1}$ is the inverse CDF (quantile function). Since in a Steiner system the maximum number of users is $N=C$, the equivalent orthogonal allocation requires a frame of length $M_{orth}=CK$.
For the sake of readability, we plot the spectral efficiency curve of the orthogonal system only for one representative case for each $K=3,4,5$ (dashed curve) and for the rest simply mark the point at which $bN \cdot R/M$ intersects with $bN \cdot R_{orth}/M_{orth}$.

In Fig.~\ref{fig:bn_combined_col} the results corresponding to the collision model with SIC are shown. There are several notable differences. 
First, the maximum rate does not exhibit such a smooth degradation as in the Full MRC case. Instead, it stays very high and close to its absolute maximum (i.e. $\log_2 (F^{-1}_{gam}(\epsilon_{tar};K,\theta) + 1)$ as in the orthogonal allocation) and then goes abruptly to $0$ once it reaches certain cut-off traffic intensity.
The maximum supportable $bN$ of a Steiner system is a non-trivial function of the order and number of its stopping sets, frame length $M$, and number of patterns $C$.
While the general trend is preserved, i.e. higher $K$ and $M$ lead to higher rates, there are some exceptions. Comparing $S(2,3,25)$ and $S(2,3,33)$, one can see that the case with $M=33$ actually performs worse. This is tied to the particularly high number of stopping sets (see Table~\ref{tab:params}). 
On the other hand, $S(2,4,37)$ can sustain higher traffic intensity than $S(2,5,41)$ despite shorter frame length. 
In this case, the reason lies in the lower $K$ and, consequently, higher number of patterns ($111$ vs $82$).
Even though the order and number of stopping sets is similar, the probability of their occurrence is effectively lower in $S(2,4,37)$.

\section{Conclusions}\label{sec:concl}
In this work we have proposed and investigated the usage of deterministic access patterns to provide ultra-reliable communication for a group of intermittently active users sharing a pool of resources. The patterns, which are a realization of a Steiner system, aim to control the number of collisions and interference among users. 
This feature leads to significant gains in terms of outage probability compared to an approach were the choice of channel resources is fully random.
In our evaluations we have considered two different signal models - based on destructive collisions and Full MRC, and two receiver processing techniques - with and without SIC.
In this work we have also developed simple approximations for the outage probability in a collision model with SIC and Full MRC model without SIC that closely match the simulation results. Such approximations are particularly important in the context of ultra-reliable systems where the number of required samples/simulations needed to properly assess the performance is often infeasible.

\appendices

\section{Proof of Proposition \ref{proposition:free:random}}\label{sec:appendix:proposition1}
\begin{proof}
 Let us  define a probability space where the \emph{samples}  are different ways the   $U-1$ other users can select the locations of their packets inside the frame. We will call these samples configurations.
 An \emph{event} is then a set of configurations, and has a certain probability.

Let us now fix $K'$ of the $K$ slots of a given user $u$. 
Denote with $A_i$ an event that consists of all the configurations comprising the fixed set $K'$ and at least one of the remaining $K-K'$ slots, denoted by $i$, 
are free of interference. 
This gives us  $K-K'$ different sets  $A_i$.

Given now an $n$-element subset $J$ of  $\{1, \dots, K-K' \}$, then  the probability for an event $\bigcap_{j\in J} A_j$ to appear is $P(\bigcap_{j\in J} A_j)=\left(\binom{M-K'-n}{K}/\binom{M}{K}\right)^{U-1}=T_n$, independent of the selected $J$ and $K'$. This can be directly seen as $\bigcap_{j\in J} A_j$ consists of all the configurations that leave at least fixed $n+K'$ slots of user $u$ interference free.  

 Let us  denote with $\mathcal{S}$ the set consisting of all the configurations where at least the fixed $K'$ slots are interference  free. By the complementary form of the inclusion-exclusion principle we then have that
\begin{equation}\label{eq:inclusion}
P\left( \mathcal{S} \setminus \bigcup_{i=1}^{K-K'} A_i\right)  =\sum_{n=0}^{K-K'} (-1)^{n}\binom{K-K'}{n} T_n
\end{equation}
where $T_0=P(\mathcal{S})=\left(\binom{M-K'}{K}/\binom{M}{K}\right)^{U-1}$.
Here the set $\bigcup_{i=1}^{K-K'} A_i$ is an event that contains all the configurations that leave the fixed $K'$ slots and at least one other of the slots occupied by the user $u$ free of interference. Then the set $\mathcal{S} \setminus \bigcup_{i=1}^{K-K'} A_i$ is an event that consists of all the configurations, where the fixed $K'$ slots are free, but none of the other $K-K'$ occupied by the user  $u$. In other words, it is the set of those configurations, where exactly $K'$ of $K$ slots are free.  One can place $K'$ packets to $K$ available slots in $\binom{K}{K'}$ different ways. Hence the final result is obtained by multiplying  \eqref{eq:inclusion} with  the term $\binom{K}{K'}$.
\end{proof}

\section{Proof of Proposition \ref{proposition:free:steiner}}\label{sec:appendix:proposition2}
\begin{proof}
  The probability that a specific set of $K'$ slots  selected by the user $u$ is not occupied by the packets of remaining $U-1$ users is given by
$$
\left(\frac{\binom{C-1-K'(D-1)}{U-1}}{\binom{C-1}{U-1}}\right)
$$
because there are $C$ patterns in total (including pattern of user $u$) and $K'(D-1)$
of them share a slot  with the $K'$ slot set of pattern of user $u$.
The proof follows that of Proposition~\ref{proposition:free:random} verbatim, after realizing that now
$$
P \left( \bigcap_{j\in J} A_j \right)=\left(\frac{\binom{C-1-(D-1)(n+K')}{U-1}}{\binom{C-1}{U-1}}\right)
$$
for any $n$ elements set of $J$.
\end{proof}

\section{}\label{sec:appendix:}
\begin{figure}[H]	
	\centering
	\includegraphics[width=0.8\linewidth]{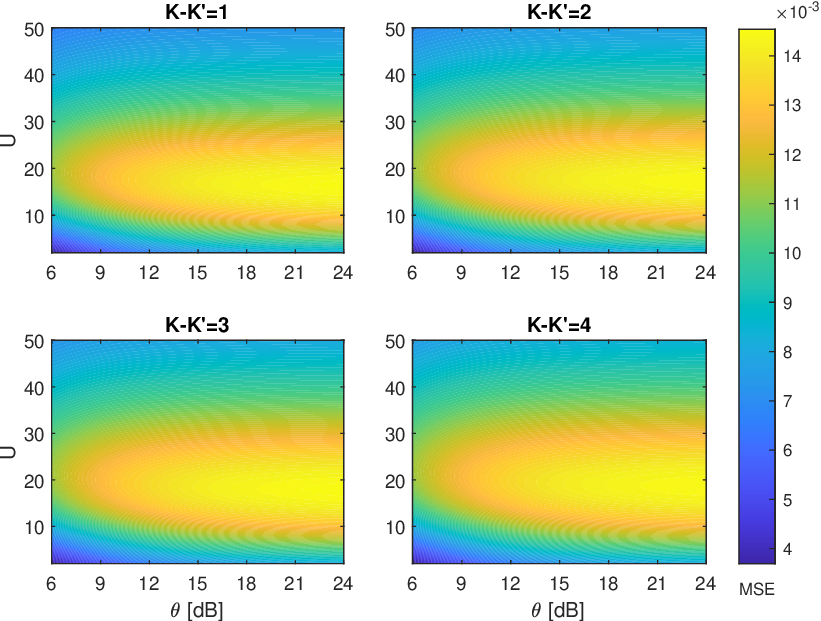}
	\caption{Results of least-squares approximation of \eqref{eq:i_sinr} with gamma distribution.}
	\label{fig:heatmap_mse}
\end{figure}

A relatively good approximation for the expression \eqref{eq:tot_sinr} can be obtained by first approximating \eqref{eq:i_sinr} with a gamma distribution, i.e. finding $f_{\text{gam}}(z;k_{K'},\alpha_{K'}) \approx f_{I-SINR}\big(z;\theta|U,K'\big)$. This can be done e.g. by solving the following optimization problem to find the suitable coefficients:

\begin{argmini}|s|[2]           	
    {k_{K'},\alpha_{K'}}           	
    { \!\! \int_{A} \! \big[ f_{I-SINR}\big(z;\theta|U,K'\big) \!-\! f_{\text{gam}}\big(z;k_{K'},\alpha_{K'}\big) \big]^2 dz 
    }   
    {\label{eq:opt1}}   	
    {}                                
    \addConstraint{\!k_{K'} > 0,\quad}{\alpha_{K'} > 0 
    }    
\end{argmini}
where $A=[0,2^R-1]$, since we are interested in the outage probability and hence concerned with a good fit only in that region. 
In Fig. \ref{fig:heatmap_mse} we show as an example the results of such fitting for the Random selection in the SINR range $[0, 2^{2}-1]$.
Goodness of the fit tends to be the lowest for a medium number of interferers which is also reflected in the final approximation in Fig.~\ref{fig:capture_single}, as it becomes less tight the higher the traffic intensity $bN$.

Since the SINRs of individual interfered slots are i.i.d and for a given $K'$ each is now represented by a gamma distribution with parameters $k_{K'}$, $\alpha_{K'}$, we have simply that 
\begin{equation}
    f_{\text{gam}}(z;k_{K'},\alpha_{K'})^{\ast (K-K')} = f_{\text{gam}}(z;(K-K')k_{K'},\alpha_{K'})
    \label{eq:gam_conv}
\end{equation}
Lastly, to combine \eqref{eq:gam_conv} with the contribution from the interference-free slots $f_{\text{gam}}(z;K',\theta)$,
we can use the result in \cite{gamma_sum} which provides the analytical expression for the sum of two gamma distributed RVs with arbitrary parameters:
\begin{equation}
\begin{split}
 & \tilde{f}_{tot,SINR}\big(z;\theta|U\big) \\
 &\;\; =\!\! \sum_{K'=0}^{K} \! p(K'|U) f_{\text{gam}}\big(z;K',\theta\big) \ast f_{\text{gam}}\big(z;(K-K')k_{K'},\alpha_{K'}\big) \\
 &\;\; =\!\! \sum_{K'=0}^{K} \! p(K'|U) \Big( \frac{\alpha_{K'}}{\theta} \Big)^{\!K'} \! \frac{ z^{\kappa-1} e^{-\frac{z}{\alpha_{K'}}}}{\alpha_{K'}^{\kappa}\Gamma(\kappa)} \tensor[_1]{F}{_1} \!\! \Big( K'; \kappa;  
 z\frac{\theta -\alpha_{K'}}{\alpha_{K'}\theta} \Big)
\end{split}
\end{equation}
where $\kappa=K'+(K-K')k_{K'}$ and $\tensor[_1]{F}{_1} (\cdot;\cdot;\cdot)$ is a Kummer’s confluent hypergeometric function. 
The CDF in this case becomes
\begin{equation}
\begin{split}
 & \tilde{F}_{tot,SINR}\big(z;\theta|U\big) \\
 & \;\; = \!\! \sum_{K'=0}^{K} \! p(K'|U) F_{\text{gam}}\big(z;K',\theta\big) \! \ast f_{\text{gam}}\big(z;(K-K')k_{K'},\alpha_{K'}\big) \\
 & \;\; = \!\! \sum_{K'=0}^{K} \! p(K'|U) \left( \frac{\alpha_{K'}}{\theta} \right)^{K'}  \! \tensor[_{[\frac{z}{\alpha_{K'}}]2}]{F}{_1} \!\left( \kappa, K'; \kappa;  \left( 1 - \frac{\alpha_{K'}}{\theta} \right) \right)
\end{split}
\end{equation}
where $\tensor[_{[\cdot]2}]{F}{_1} ( \cdot, \cdot; \cdot;  \cdot)$ is the incomplete Gauss hypergeometric function.

\ifCLASSOPTIONcaptionsoff
  \newpage
\fi



%

\bibliographystyle{IEEEtran}
\bibliography{Biblio}




\end{document}